\documentclass[doublecol]{epl2} 
\usepackage{amssymb}   
\usepackage{amsmath}

\title{Controlled engineering of spin polarized transport properties in a zigzag
  graphene nanojunction} \shorttitle{Controlled engineering of spin
  polarized transport properties} 

\author{Sudin Ganguly\inst{1} \and Saurabh Basu\inst{1} \and Santanu K. Maiti\inst{2}}
\shortauthor{S. Ganguly \etal}

\institute{ \inst{1} Department of Physics, Indian Institute of
  Technology Guwahati, Guwahati-781039, Assam, India
  
  \inst{2} Physics and Applied Mathematics Unit, Indian Statistical
Institute, 203 Barrackpore Trunk Road, Kolkata-700 108, India
}


\abstract{ We investigate a novel way to manipulate the spin polarized
  transmission in a two terminal zigzag graphene nanoribbon in
  presence of Rashba spin-orbit (SO) interaction with circular shaped
  cavity engraved into it. A usual technique to control the spin
  polarized transport behaviour of a nanoribbon can be achieved by
  tuning the strength of the SO coupling, while we show that an
  efficient engineering of the spin polarized transport properties can
  also be done via cavities of different radii engraved in the
  nanoribbon. Simplicity of the technique in creating such cavities in
  the experiments renders an additional handle to explore transport
  properties as a function of the location of the cavity in the
  nanoribbon. Further, a systematic assessment of the interplay of the
  Rashba interaction and the dimensions of the nanoribbon is
  presented. These results should provide useful input to the
  spintronic behaviour of such devices. In addition to the spin
  polarization, we have also included an interesting discussion on the
  charge transmission properties of the nanoribbon, where, in absence
  of any SO interaction a metal-insulator transition induced by the
  presence of a cavity is observed.}

\begin{document}

\maketitle

\section{\label{sec1}Introduction}
Spin-based electronics or spintronics is one of the most promising
field for future power-consuming high operating speed, new forms of
information storage and logic devices~\cite{wolf}. The key factor for
the development of spin-based electronics~\cite{zutic} is the
fine control of the spin-polarized current. For this purpose, a major
challenge is developing a suitable spin transport channel with long
spin diffusion length and spin lifetime. Graphene~\cite{novo},
believed to be a very promising candidate in the spintronic
applications, owing to the achievement of room-temperature spin
transport with long spin-diffusion lengths (up to $\sim
100\,\mu$m)~\cite{luis,tombros, zomer,yang,han}. Graphene has also
several interesting electronic and transport properties~\cite{neto},
that make it very attractive for spintronic applications, such as
quasirelativistic band structure~\cite{novo,zhang}, unconventional
quantum Hall effect~\cite{novo,zhang,vp}, half
metallicity~\cite{jun,lin} and high carrier
mobility~\cite{du,bolotin}.

Recent experimental realization of freestanding graphene nanoribbons 
(GNRs)~\cite{meyer,moro} has generated renewed interest in carbon-based 
materials with exotic properties. GNRs are basically single strips of 
graphene where electronic properties~\cite{fujita,saito} depend on the 
geometry of the edges and the lateral width of the nanoribbons~\cite{nakada}. 
Depending upon the edge structures, GNRs can be of two types, namely 
armchair graphene nanoribbon (AGNR) and zigzag graphene nanoribbon (ZGNR). 
Irrespective of the width of the ZGNR, they are always metallic with zero 
bandgap, while the AGNRs are conditionally metallic, that is when the 
lateral width $N_y = 3M-1$ ($M$ is an integer), else the AGNRs are 
semiconducting in nature~\cite{fujita} with a finite band gap. GNR is 
also known to have long spin-diffusion length, spin relaxation time, 
and electron spin coherence time~\cite{yazyev-prb,yazyev-prl,cantele}.

SO coupling (SOC) plays a crucial role in spintronic devices. 
Two kinds of SOC can be present in graphene, the intrinsic SOC and the 
Rashba SOC (RSOC)~\cite{km1,km2}. The strength of the intrinsic SOC is 
negligibly small in pristine graphene 
(up to $\sim$ 0.01-0.05 meV)~\cite{yao-prb,jc-prb}. On the other hand, 
the strength of the Rashba SOC can be modulated by an external electric 
field or by using substrates. Recent observations showed that the 
strength of the Rashba SOC can be enhanced up to 100 meV from Gold (Au) 
intercalation at the graphene-Ni interface~\cite{marchenko}. A Rashba 
splitting about 225 meV in epitaxial graphene layers grown on the surface 
of Ni~\cite{dedcov} and a giant Rashba SOC ($\sim$ 600 meV) from Pb 
intercalation at the graphene-Ir surface~\cite{calleja} are noted in 
experiments. In view of the above discussion, GNRs can be used as 
spin-based devices. Consequently a variety of graphene-based spintronic 
devices have been proposed~\cite{frank,zeng,kim,jozsa,y-t,bennett,
cai,chico,qzhang}, for example, prediction of spin-valve devices based 
on graphene nanoribbons exhibit giant magnetoresistance (GMR)~\cite{kim}, 
spin-valve experiment on GNR~\cite{jozsa}, study of spin polarization 
and giant magnetoresistance in GNR~\cite{y-t}, experiments of GNR as 
field-effect transistor~\cite{bennett} and p-n junctions~\cite{cai} 
using bottom-up fabrication technique and many more~\cite{chico,qzhang}. 
However, most of the studies have been dedicated to making GNR as 
information storage or attempted to get larger spin polarization.  For 
a complete realization of spintronic applications, a fine control 
of the spin polarization is highly desirable.

Further, a few of the studies have been dedicated on the transport
properties and band structures of graphene nanoribbon in presence of
antidots~\cite{lv,ajm,rosa,mads}, which, in essence, is same as our
`cavity'.  These studies were basically searching for possible ways to
open up energy gaps at the Dirac points to propose an on-off switching
mechanism using graphene nanoribbons. In this work, in addition to the
charge transport properties being alluded briefly, our central focus
is to deliberate on the spin polarized transport in order to explore
spintronic application elaborated in the subsequent discussion.

In this letter, we have proposed a possible way of tuning the spin
polarization in ZGNR modulated with a circular shaped cavity in
presence of Rashba SOC. We have shown that the spin polarized
transmission of a two terminal ZGNR can be tuned by varying the radius
of the circular cavity. Recent experimental advancement in making
nanoscale devices also indicate that our proposed theoretical model
can be achieved experimentally. Michael {\it et al.} showed that it is
possible to realize nanometer-scale pores in graphene by controlled
exposure to the focused electron beam of a transmission electron
microscope~\cite{michael} and most importantly they do not evolve over
time. Nanolithographic technique~\cite{zhang-nano}, template
growth~\cite{nathan} technique can be used to achieve desired shape
and size of nanopores with great accuracy.

If graphene is grown on a corrugated target substrate, it
  acquires the morphology of the substrate, leading to
  ripples~\cite{michael}. This, in turn, enhances local SO
  coupling~\cite{neto-prl}. For example a $\sim$10 meV SO splitting has
  been reported in flat Au monolayer~\cite{vary}, whereas graphene-Au
  hybridization enhances the Rashba splitting upto 100
  meV~\cite{marchenko}. Moreover, intercalation of a Pb monolayer
  between graphene and Ir(111) results into an almost negligible
  structural corrugation and reduces the overall SO
  coupling~\cite{otrokov}. In this work, all the calculations are
  carried out by assuming a flat GNR (uncorrugated) and thus such
  corrugated scenario is not taken under consideration.

We organize our paper as follows. In the following section, we present
the proposed model and the theoretical formalism for the total
transmission and spin polarized transmission using the Green's
function technique.  Subsequently, we include an elaborate discussion
of the results where we have demonstrated cavity effect in ZGNR on the
spin polarization, that is, how the spin polarized transmission
behaves with the dimension of the system and with the position of the
centre of the cavity. We end with a brief summary of our results.
\section{\label{sec2}Model and Theoretical formulation}

\begin{figure}[ht]
{\centering 
\resizebox*{7cm}{3.5cm}{\includegraphics{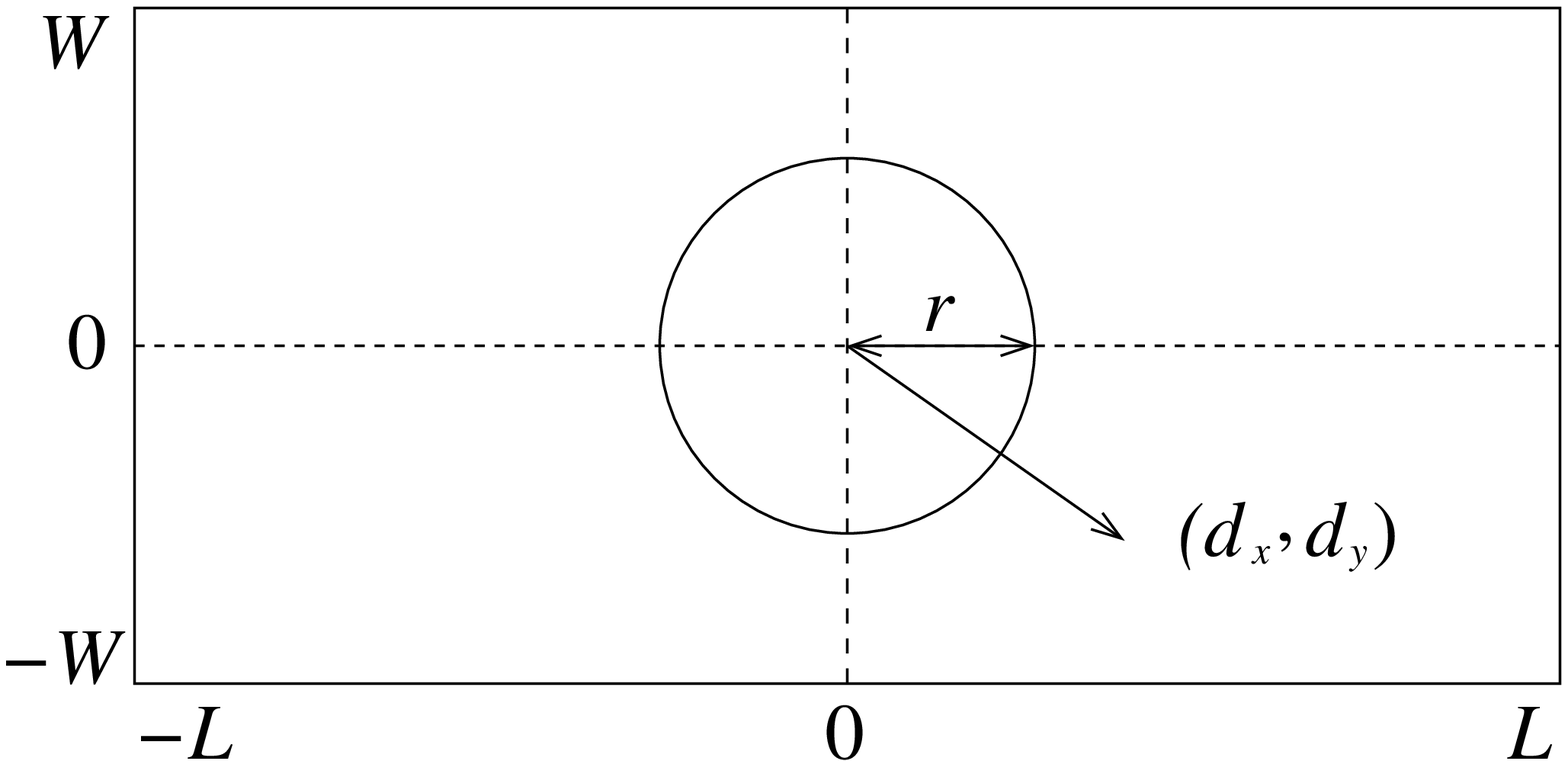}}\\
\resizebox*{7cm}{3.5cm}{\includegraphics{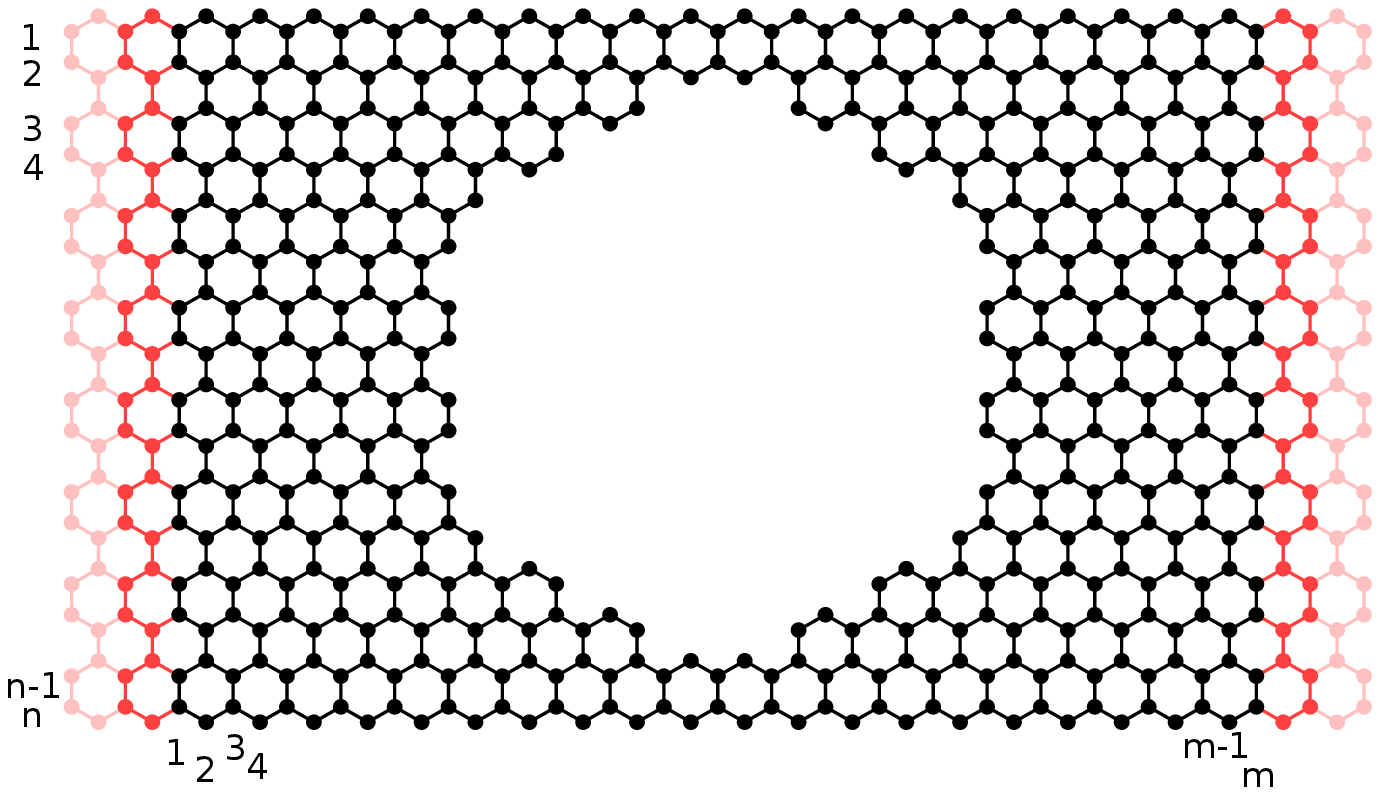}}
\par}
\caption{(Color online). (Top) Schematic diagram for a zigzag graphene
  nanoribbon with a circular-shaped single cavity. $2L$ and $2W$ are
  the length and the width of the system respectively. The radius of
  the cavity is denoted by $r$ and the postion coordinate of the
  centre of the cavity is $(d_x,d_y)$. (Bottom) A real setup for
  calculating the spintronic properties.}
\label{single_cavity}
\end{figure}

We consider a zigzag graphene nanoribbon (ZGNR) inscribed with a
circular-shaped cavity structure as shown in
Fig.~\ref{single_cavity}(top).
The length and the width of the ZGNR are $2L$ and $2W$
respectively. The radius of the circular-shaped cavity is $r$ and the
position of the centre of the cavity is $(d_x,d_y)$. We consider a
two-terminal setup as shown in Fig.~\ref{single_cavity}(bottom) in
order to calculate the charge and spintronic properties of the ZGNR,
where two leads are attached to the central scattering region (ZGNR in
this case). The leads are denoted by red colour and are semi-infinite
and free from any kind of SO interactions. The Rashba
interaction is assumed to be present in the central scattering region
only.

The length and the width of the system can be defined in the following
way. Since along the length, the system has a zigzag shape and along
the width an armchair, we call the system as mZ-nA as shown in
Fig.~\ref{single_cavity}(bottom). The system dimension can
  also be expressed in nm with the help of the following relations~\cite{sudin-sm}, $
  L_x = \frac{\sqrt{3}}{2}(m-1)a,\quad L_y =
  \left(\frac{3}{2}n-1\right)a $, where $a=0.142$ nm. Though the nm
  unit is useful to measure sample dimensions experimentally, the mZ-nA 
  notation is also helpful in some cases to visualize the system. Thus,
  throughout this work, we shall use nm unit as well as the mZ-nA
  notation. However, the radius of the cavity, $r$, is measured in nm.

The tight-binding (TB) Hamiltonian modelled on a ZGNR in presence of Rashba
SO interaction can be written as~\cite{km1,km2},
\begin{equation}
H= - t\sum\limits_{\langle ij\rangle}c_i^{\dagger} c_j +
i\alpha\sum\limits_{\langle ij\rangle}c_i^{\dagger} \left(
\vec{\sigma} \times {\bf\hat{d}}_{ij}\right)_z c_j
\label{h2}
\end{equation} 
where $c_i^{\dagger}=\left(c_{i\uparrow}^{\dagger} \quad
c_{i\downarrow}^{\dagger}\right)$. $c_{i\sigma}^{\dagger}$
$(\sigma=\uparrow,\downarrow)$ is the creation operator of an electron
at site $i$ with spin $\sigma$. The first term is the nearest-neighbor 
hopping term, with a hopping strength $t=2.7$ eV. The second term is the 
nearest-neighbor Rashba term which explicitly violates $z\rightarrow -z$ 
symmetry. $\vec{\sigma}$ denotes the Pauli spin matrices and $\alpha$ 
is the strength of the Rashba SO interaction. ${\bf\hat{d}}_{ij}$ is 
the unit vector that connects the nearest-neighbor sites $i$ and $j$.

The total transmission coefficient $T$ can be calculated
via~\cite{caroli,Fisher-Lee,dutta},
\begin{equation}
T = \text{Tr}\left[\Gamma_L {\cal G}_R
  \Gamma_R {\cal G}_A\right]
\end{equation}
${\cal G}_{R(A)}$ is the retarded (advance) Green's function. 
$\Gamma_{L(R)}$ are the coupling matrices representing the coupling between 
the central region and the left (right) lead.

Finally, we calculate spin polarized transmission coefficient $P_s$
from the relation as~\cite{chang},
\begin{equation}
P_s = \text{Tr}\left[\hat{\sigma}_s\Gamma_L G_R
  \Gamma_R G_A\right]
\end{equation}
where, $s=x,y,z$ and $\sigma$ denote the Pauli matrices.

We have studied the charge and spin properties of ZGNR modulated with
a circular-shaped cavity structure.

\section{\label{sec3}Results and Discussions}

We set the hopping term $t=2.7$ eV. All the energies are measured in
units of $t$. For most of our numerical calculations, we have used
KWANT~\cite{kwant}. In order to visualize the critical roles
of RSOI, and, for the sake of comprehensive analysis we have taken 
the RSOI strength over a wide range varying from a moderate value
($0.05t \sim 0.135$ eV) to a very large one ($0.5t\sim 1.35$ eV), which is referred to as a
giant Rashba SO coupling.

Before going into the results, there is one important point about the
spin polarized transmission that should be mentioned. Since the ZGNR
has longitudinal mirror symmetry along $x$ and $z$ axes, the $x$ and
$z$ components of the spin polarized transmission are zero. Only the
$y$ component has a non-zero value (due to the finite width of the
ZGNR along $y$ direction)~\cite{chico,qzhang}. Hence from now on, we
shall call the $y$-component of the spin-polarized transmission
coefficient as $P$.

\begin{figure}[ht]
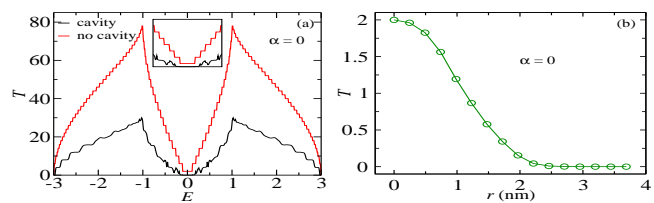

{\hfill
\includegraphics[height=0.14\textwidth,
  width=0.23\textwidth]{fig2a.eps}\hfill
\includegraphics[height=0.14\textwidth,
  width=0.23\textwidth]{fig2b.eps}
\hfill}
\caption{(Color online). (a) The total transmission coefficient $T$
  is plotted as a function of the Fermi energy $E$ in the clean
  limit. The system dimension is 9.8 nm-8.4 nm (81Z-40A). The red
  colour denotes the case without cavity, and, the black one with
  the cavity where the radius of the cavity is fixed at $r\sim 2.5$ nm. 
  (b) Dependence of $T$ on $r$ setting the Fermi energy $E=0$.}
\label{clean}
\end{figure}

In Fig.~\ref{clean}, the total transmission coefficient $T$ is
plotted as a function of the Fermi energy $E$ in the absence of the
Rashba SO interaction. We set the dimension of the ZGNR as 9.8 nm-8.4
nm (81Z-40A). The red colour denotes the plot without a cavity,
while the black one denotes the one with cavity with radius $r\sim
2.5$ nm.
When the cavity is absent in the system, $T$ shows usual discrete step-like
feature emphasizing the occurrence of quantum transport at discrete energy
values. As the cavity is introduced, the transmission loses its step-like
feature and also the magnitude of $T$ gets suppressed. Another important
observation is that, in the absence of the cavity a $2e^2/h$ plateau is
observed around the zero of the Fermi energy. This $2e^2/h$ plateau plays a
vital role in determining the presence of edge
states~\cite{fujita,sudin-mrx}. However, this $2e^2/h$ conductance
plateau vanishes as we introduce cavity in the system. For better
clarity the behaviour of $T$ is plotted, near the zero of the Fermi energy,
in the inset in Fig.~\ref{clean}(a). In order to have more information
about the vanishing nature of the $2e^2/h$ plateau near the zero
of the Fermi energy, we have plotted the total transmission $T$ as a
function of the radius of the cavity as shown in Fig.~\ref{clean}(b).
Here, we set the Fermi energy at zero. From the Fig.~\ref{clean}(b),
it is observed that $T$ smoothly falls off as we increase the radius
of the cavity and vanishes completely after a certain value of $r$.
For $r\lesssim 2.5$ nm, the system shows metallic behavior, and, for
$r\gtrsim 2.5$ nm it acts as an insulator. Thus, {\em a metal-to-insulator
(MI) transition across the zero Fermi energy can be achieved in a ZGNR
using a cavity inscribed.}
Most importantly, this feature is observed in the absence of any kind
of SO interaction. Basically, as we increase the radius of the cavity,
the transmission gap is enlarged due to the finite-size effect and as a
result the MI transition occurs in the system. The same findings for
\begin{figure*}[ht]
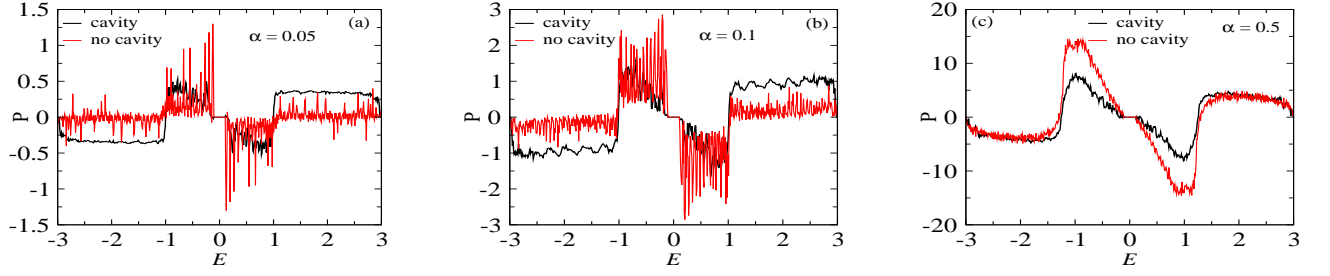

{\hfill
\resizebox*{5.0cm}{3.5cm}{\includegraphics{fig3a.eps}}\hfill
\resizebox*{5.0cm}{3.5cm}{\includegraphics{fig3b.eps}}\hfill
\resizebox*{5.0cm}{3.5cm}{\includegraphics{fig3c.eps}}\hfill
\par}
\caption{(Color online). Spin polarized transmission as a function of the 
Fermi energy $E$ for three different strengths of RSOI ($\alpha$) considering
the system dimension 9.8 nm-8.4 nm (81Z-40A), where the black and red colored curves
correspond to the systems with and without cavity respectively. The radius of the cavity 
is fixed at $r\sim 2.5$ nm.}
\label{sp_e}
\end{figure*}
\begin{figure*}[ht]
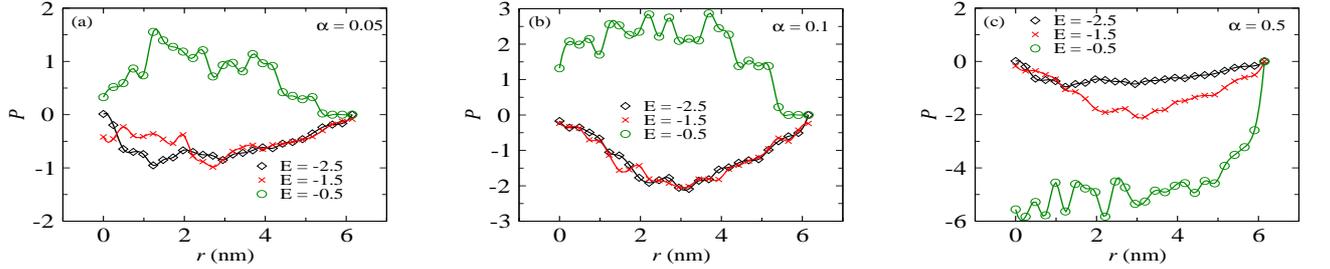

{\hfill
\resizebox*{5.0cm}{3.5cm}{\includegraphics{fig4a.eps}}\hfill
\resizebox*{5.0cm}{3.5cm}{\includegraphics{fig4b.eps}}\hfill
\resizebox*{5.0cm}{3.5cm}{\includegraphics{fig4c.eps}}\hfill
\par}
\caption{(Color online). $P$-$r$ characteristics at three typical energis 
considering the system size $\sim$ 15 nm-12.6 nm (121Z-60A).}
\label{r_var}
\end{figure*}
larger system dimensions have also been done and we have found that though 
the behaviour of $T$ is not exactly the same as shown in Fig.~\ref{clean}(b) 
but in all the cases, beyond a certain value of $r$, the system looses 
the plateau and becomes an insulator.

Let us now include the effect of the Rashba SO interaction and 
explore the
interplay between RSOI and cavity on the spin polarized transport. The spin 
polarized transmission behaviour in the absence and presence of cavity are 
presented in Fig.~\ref{sp_e}. All the plots for the spin polarized 
transmission are antisymmetric in nature as a function of the Fermi energy 
$E$ which is due to the electron-hole symmetry of the system. The radius 
of the cavity is taken as $r\sim 2.5$ nm and the system size is kept fixed 
at 9.8 nm-8.4 nm (81Z-40A). For smaller values of RSOI, namely, $\alpha=0.05$ 
and $0.1$, the spin polarized transmission shows reasonably large 
fluctuations for the cavity-free system as a function of $E$, especially within
the range $-1\leq E \leq 1$ (red lines of Figs.~\ref{sp_e}(a) and (b)). 
Whereas the fluctuations get suppressed in the presence of the cavity
(black lines of Figs.~\ref{sp_e}(a) and (b)). Interestingly we see that for 
these smaller values of RSOI, much larger spin polarization is achieved in
presence of cavity beyond the range $E \geq 1$ and $E \leq 1$, {\em which may 
give an important hint of achieving a high degree of cavity-induced spin 
polarization, and, to the best of our knowledge, this phenomenon has not been
discussed so far in literature.} Along with the cavity, the fluctuations get
also suppressed with increasing the strength of RSOI, as clearly seen from
Fig.~\ref{sp_e}(c). Though a smooth curve is obtained, which is always 
favorable in the context of a physical measurement, but the strength of RSOI 
in this case (Fig.~\ref{sp_e}(c)) is too large and has not been achieved 
experimentally so far. For our theoretical study, we choose one such large 
value of $\alpha$ along with other smaller ones, for the sake of a complete 
analysis. Comparing the spectra given in Fig.~\ref{sp_e}, one can see that 
the overall magnitude of $P$ enhances with the rise of $\alpha$.

The above analysis raises an obvious question whether one can 
tune spin 
polarized transmission by varying the radius of the cavity. To explore it,
let us look into the spectra given in Fig.~\ref{r_var} where $P$-$r$
characteristics are shown at three typical energies ($E=-2.5$, $-1.5$ and
$-0.5$ (in units of $t$)) considering a reasonably large system, compared
to the previous one, having system size 15 nm-12.6 nm (121Z-60A). We
choose arbitrarily these typical energies where transmission probability $T$
is sufficiently high ($E=0$ is not be a good choice, as in this case
$T$ goes to zero beyond a critical value of $r$, as clearly seen from
Fig.~\ref{clean}). Several interesting features are noticed. For different 
values of $E$, the signs of $P$ are different depending on the dominating
spin band (up or down). The most notable aspect is that, as we increase $r$, 
up to $r\sim$ 3 nm, the degree of polarization increases, beyond which it 
starts to decrease. For larger values of $r$ i.e., when $r\approx 6$ nm, 
$P$ almost drops to zero. This vanishing nature of $P$ at large $r$ is
associated with the fact that the number of conducting channels gradually 
decreases with removing the lattice sites (that is, increasing cavity size), 
and beyond a critical radius, almost no paths are available for electronic 
conduction which yields vanishingly small polarization. But below this 
critical radius that is, where $P$ is finite one may always expect reduced
spin polarization with increasing $r$ as spin dependent scattering 
mechanism, associated with Rashba term, gets weaker with the removal 
of lattice sites. But the results of Fig.~\ref{r_var} show a non-monotonic
nature where $P$ initially increases for smaller $r$, and after reaching
to a maximum, it ($P$) decreases for large values of $r$. There
are several possible reasons for that as the polarization effectively
depends on which spin channel (up or down) is dominating. One possible
reason is the existence of large fluctuations at low cavity sizes, where 
for a particular energy $E$, $P$ differs by reasonably large values for two distinct 
cavity sizes. This difference gradually decreases with large values of
$r$, consistent with the suppression of fluctuations, and in that
case, reduced values of $P$ is observed with $r$, as expected.

\begin{figure}[ht]
{\centering \resizebox*{5cm}{3cm}{\includegraphics{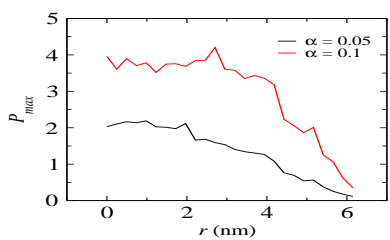}}\par}
\caption{(Color online). $P_{max}$-$r$ characteristics for two different
strengths of RSOI considering the same system size as taken in 
Fig.~\ref{r_var}.} 
\label{p_max}
\end{figure}

From the spectra given in Figs.~\ref{sp_e} and \ref{r_var} we get a
clear hint that one can tune $P$ by regulating the cavity size, which
is our primary goal of the present analysis. Now, as the value of $P$
depends on the specific choice of the energy $E$, we need to scrutinize 
further by considering wider range of energy in order to understand the precise role
of cavity on spin polarization. In order to do that, in Fig.~\ref{p_max} 
we show the variation of $P_{max}$ (maximum value of $P$) with $r$ for 
two different strengths of RSOI (namely $\alpha=0.05$ and 0.1) considering the same system size as taken 
in Fig.~\ref{r_var}. $P_{max}$ is computed by taking the maximum value of
$P$ over the complete energy window $[-3:3]$ (in units of $t$). It is
seen that $P_{max}$ decreases with $r$ providing minor fluctuations,
and eventually reduces nearly to zero. The decreasing nature can easily be
understood, following the results of Figs.~\ref{sp_e} and \ref{r_var}, 
as the scanning is done over the full energy window. The reduction of 
$P_{max}$ with $r$ does not actually yield a negative response towards 
tuning mechanism using cavity, as in the calculation of $P_{max}$ it
takes the maximum value of $P$. While in an experimental situation whenever we
 talk about possible tuning of $P$ with cavity, we need to fix
the Fermi energy at a particular value and there is a high possibility 
to achieve favorable spin polarization and even much higher polarization
than the cavity-free case, as reflected in Figs.~\ref{sp_e} 
and \ref{r_var}.

\begin{figure}[ht]
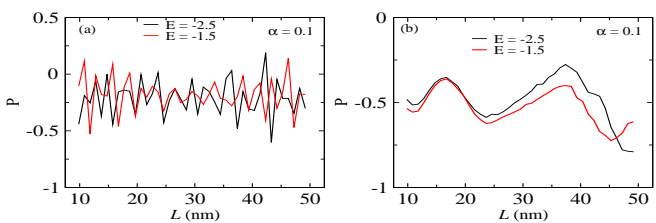

{\centering
\resizebox*{4.2cm}{2.8cm}{\includegraphics{fig6a.eps}}
\resizebox*{4.2cm}{2.8cm}{\includegraphics{fig6b.eps}}
\par}
\caption{(Color online). $P$ as a function of the length of the system $L$,
considering width of the system as $\sim 16.9$ nm ($80A$) and the RSOI 
strength $\alpha=0.1$. (a) Without cavity, and (b) with cavity having 
$r\sim 3.7$ nm.}
\label{l_var}
\end{figure}

\begin{figure}[ht]
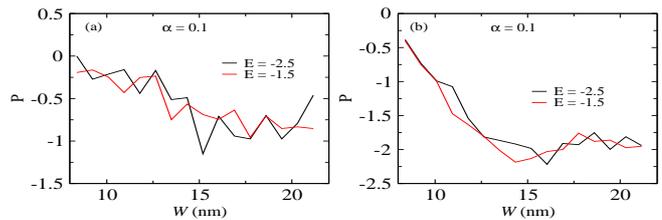

{\centering
\resizebox*{4.2cm}{2.8cm}{\includegraphics{fig7a.eps}}
\resizebox*{4.2cm}{2.8cm}{\includegraphics{fig7b.eps}}
\par}
\caption{(Color online). $P$ as a function of the width of the system $W$,
taking the length of the system as $\sim 14.8$ nm (121Z) and the RSOI 
strength $\alpha=0.1$. (a) Without cavity, and (b) with cavity having
$r\sim 3.7$ nm.}
\label{w_var}
\end{figure}

For completeness, we have also studied the finite size effects on the
behaviour of spin polarized transmission, as shown in Fig.~\ref{l_var}
and Fig.~\ref{w_var} respectively.
In Fig.~\ref{l_var}(a), we have shown the effect of the length (in
units nm) of the system on $P$ without cavity and in
Fig.~\ref{l_var}(b) with cavity of radius $r\sim 3.7$ nm. The width of
the system is taken $\sim 16.9$ nm ($W=80A$). We set $\alpha=0.1$. The
spin polarized transmission shows large fluctuations in
Fig.~\ref{l_var}(a). However, in presence of the cavity, the behaviour
of $P$ is smooth and oscillating in nature as a function of the length
of the system.  Further, the effect of the width $W$ (in units of nm)
of the system on the spin polarized transmission is shown is
Fig.~\ref{w_var}. Here we have taken the system length as $\sim 14.8$
nm (121Z) and fixed the Rashba SO interaction strength at $\alpha=0.1$. 
In the absence of cavity, $P$ shows more fluctuations (Fig.~\ref{w_var}(a)) 
than that with the cavity case (Fig.~\ref{w_var}(b). In both the cases, 
$P$ initially increases with $W$ up to $W\sim$ 15-16 nm, then it tends to 
saturate. However, the
rate of change of $P$ is more in presence of cavity compared to the
cavity free case. Now, as we increase the width of the system the
number of modes available for transport increases. As a result,
initially $P$ increases
\begin{figure*}[ht]
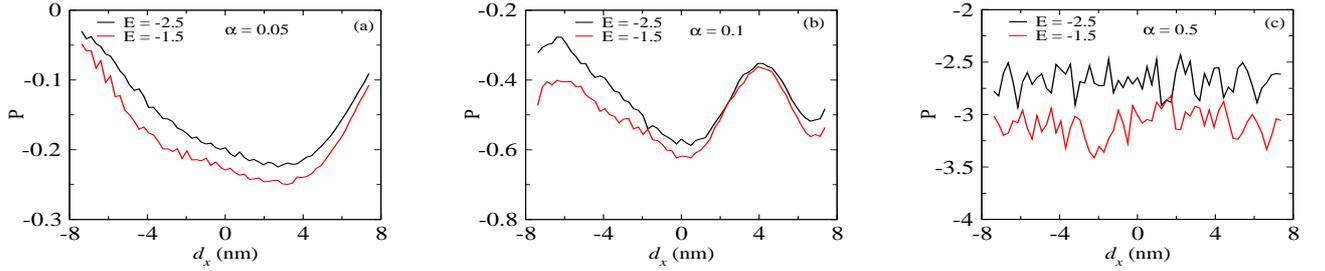

{\hfill
\resizebox*{5.0cm}{3.5cm}{\includegraphics{fig8a.eps}}\hfill
\resizebox*{5.0cm}{3.5cm}{\includegraphics{fig8b.eps}}\hfill
\resizebox*{5.0cm}{3.5cm}{\includegraphics{fig8c.eps}}\hfill
\par}
\caption{(Color online). $P$ as a function of position of the centre of 
the cavity $d_x$, considering the dimension of the system as 
$\sim$ 25 nm-8.4 nm (201Z-40A). The centre of the cavity is moving along 
$y=0$ line, and here we choose $r\sim 3.7$ nm.}
\label{dx}
\end{figure*}
with $W$. However, when $W$ is large enough compared to the radius 
of the cavity, which is the case for $r\sim 3.7$ nm, the system tends to 
behave as that of without cavity.

In all the results discussed so far the centre of the cavity was
fixed at the origin or in other words, the position of the centre of
the cavity was fixed at $\left(d_x,d_y=0,0\right)$ (see
Fig.~\ref{single_cavity}(top)). It is also important to see how the
position of the centre of the cavity affects the spin polarized
transport. In other words, in order to answer a rather important
question whether there is anything like a most suitable position for
the cavity to exist in the system that would yield the maximum spin
polarized transmission. To do so, we have considered two different
scenario. First, we have varied the $x$-coordinate $\left(d_x\right)$
keeping the $y$-coordinate fixed at $y=0$ $\left(d_y\right)$. In the
second case, we have varied $d_y$ along the $x=0$ line. Thus the
translation of the cavity in both $x$ and $y$ directions were considered.

The behaviour of the spin polarized transmission as a function of the
$x$-coordinate of the centre of the cavity $d_x$ (in units of nm)
and symmetrically with respect to the origin is shown in Fig.~\ref{dx}
for two different values of the Fermi energy, namely $E=-2.5$ and
$-1.5$ in presence of Rashba SO interaction.
\begin{figure*}[ht]
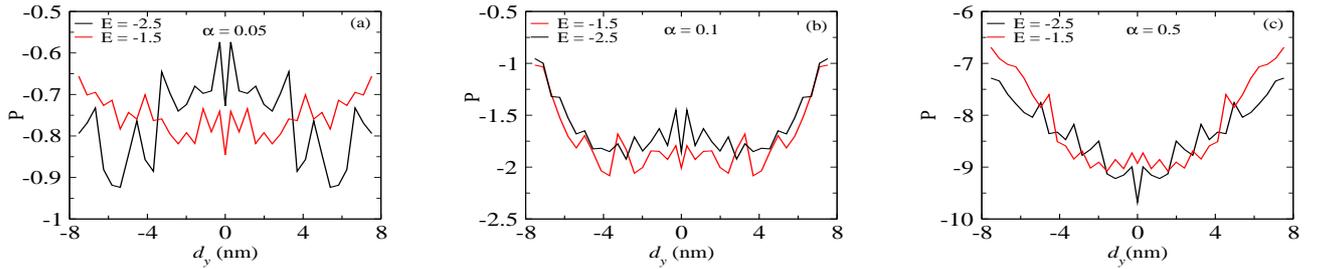

{\hfill
\resizebox*{5.0cm}{3.5cm}{\includegraphics{fig9a.eps}}\hfill
\resizebox*{5.0cm}{3.5cm}{\includegraphics{fig9b.eps}}\hfill
\resizebox*{5.0cm}{3.5cm}{\includegraphics{fig9c.eps}}\hfill
\par}
\caption{(Color online). $P$ as a function of position of the centre of 
the cavity $d_y$, taking the system size 14.8 nm-21.2 nm (121Z-100A). The 
centre of the cavity is moving along $x=0$ line, and we set $r\sim 2.5$ nm.}
\label{dy}
\end{figure*}
We have taken three different values of $\alpha$, which are
$\alpha=0.05$ (Fig.~\ref{dx}(a)), $\alpha=0.1$ (Fig.~\ref{dx}(b)) and
$\alpha=0.5$ (Fig.~\ref{dx}(c)). Here, the centre of the cavity is moving 
along $y=0$ line and the radius of the cavity is taken as $r=15$. The 
system dimension is 25 nm-8.4 nm (121Z-40A). As shown in Fig.~\ref{dx}(a), 
for $\alpha=0.05$, $P$ increases as the cavity moves from the left side 
of the system to the right and attains a maximum value around 
$d_x\approx 4$ nm. Subsequently $P$ starts to decrease as
the cavity moves towards the right end of the system. For $\alpha=0.1$
(Fig.~\ref{dx}(b)), $P$ shows little oscillatory behavior as a
function of $d_x$ and the maximum value of $P$ is observed around
$d_x=0$. For higher values of $\alpha$, that is, $\alpha=0.5$
(Fig.~\ref{dx}(c)), $P$ does not have any regular behaviour as a
function of $d_x$. This points towards possibilities of significant
interplay between the position of the cavity (in addition to the
radius as discussed above) and the strength of the Rashba interaction
with regard to its spin polarized transport properties.

The behavior of the spin polarized transmission as a function of the
$y$-coordinate of the centre of the cavity $d_y$ (in units of nm) is
shown in Fig.~\ref{dy}. Here, the cavity is translated along the $x=0$
line. The radius of the cavity is taken as $r\sim 2.5$ nm and the
dimension of the system is 14.8 nm-21.2 nm (121Z-100A). For the three
different values of $\alpha$, viz, $\alpha=0.05$ (Fig.~\ref{dy}(a)),
$\alpha=0.1$ (Fig.~\ref{dy}(b)) and $\alpha=0.5$ (Fig.~\ref{dy}(c)),
$P$ shows symmetric behaviour as a function of $d_y$ about $d_y=0$ due
to the symmetry of the position of the cavity about the $x=0$
line. The magnitude of $P$ is higher for higher values of $\alpha$,
which is obvious.

From the study of the behavior of spin polarized transmission as a
function of the positions of the centre of the cavity, it is difficult
to comment on the optimized position of the cavity, so that one can
get large spin polarized transmission. This is apparent from the fact
that there is no regular feature of $P$ as seen from Fig.~\ref{dx} and
Fig.~\ref{dy}. However, from Fig.~\ref{dy}, we can at least recommend
that, it is prudent not to put the cavity at far left or far right
edge of the system, since at these positions, $P$ has lower values
compared to other intermediate locations.

\section{\label{sec6}Conclusion}

In summary, we have studied briefly the charge transmission and more
elaborately spin polarized transport properties in a two-terminal ZGNR
modulated with a circular-shaped cavity in presence of RSOI. The charge 
transmission shows interesting feature in presence of the cavity. In 
particular, we notice a cavity-induced MI transition in the interacting 
free system. From the detailed analysis of spin polarized transport we have
shown that by varying the radius of the cavity, one can tune the spin 
polarized transmission efficiently in a two-terminal ZGNR. All the results 
are obtained for fairly large dimensions of the system. Moreover, we have 
also studied the effect of the system dimension, such as the length and 
width of the system on the spin polarized transmission. The spin polarized 
transmission oscillates with the length of the system and for larger values 
of the width it tends to saturate. However, the main focus of this work, 
that is the tuning of the spin polarized transmission, shows qualitatively 
similar behaviour for larger dimensions of the system. The position of 
the centre of the cavity is also an important parameter for tuning the 
spin polarized transmission. In order to achieve higher spin polarized 
transmission, we issue a word of caution that one should not put the cavity 
at the extreme left or at the extreme right edge of the system.

Considering the success of fabricating `holey Graphene'~\cite{hG} seamlessly, 
our work may render a new direction to engineer tunable spin polarized 
transport properties of GNRs via cavities of different shapes and sizes.

\acknowledgments SB thanks SERB, India for financial support under the
grant F. No: EMR/2015/001039.

\end{document}